| | |
|---|---|
| Title | **Chemical and Physical Effects of the Carrier Gas on the Atmospheric Pressure PECVD of Fluorinated Precursors** |
| Authors | Julie Hubert[1], Nicolas Vandencasteele[1], Jérémy Mertens[1], Pascal Viville[2], Thierry Dufour[1], Cédric Barroo[3], Thierry Visart de Bocarmé[3], Roberto Lazzaroni[2], François Reniers[1] |
| Affiliations | [1] Faculté des Sciences, Service de Chimie Analytique et de Chimie des Interfaces, Université Libre de Bruxelles, CP-255, Bld du Triomphe, B-1050 Bruxelles, Belgium<br>[2] Service de Chimie des Matériaux Nouveaux, Université de Mons- UMONS/Materia Nova, 20 Place du Parc, 7000 Mons, Belgium<br>[3] Faculté des Sciences, Chemical Physics of Materials, Université Libre de Bruxelles, CP-243, Bld du Triomphe, B-1050 Bruxelles, Belgium |
| Ref. | Plasma Processes & Polymers, 2015, Vol. 12, Issue 10, 1174-1185 |
| DOI | http://dx.doi.org/10.1002/ppap.201500025 |
| Abstract | The atmospheric pressure PECVD deposition and texturization of hydrophobic coatings using liquid fluorinated $C_6F_{12}$ and $C_6F_{14}$ precursors are investigated. The effect of the carrier gas (argon and helium) is discussed in terms of the behavior of the gas phase and of the characteristics of the deposited film. Mass spectrometry measurements indicate that the fragmentation is higher with argon while helium reacts very easily with oxygen impurities leading to the formation of $C_xF_yO_z$ compounds. These observations are consistent with the chemical composition of the films determined by XPS and the variation in the deposition rate. Moreover, the streamers present in the argon discharge affect the morphology of the surface by increasing the roughness, which leads to the increase in the hydrophobicity of the coatings. |

# 1. Introduction

Low-pressure plasma deposition of fluorocarbon films has been widely studied in the last decades.[1–6] The main goal of those works was to evaluate the capability of tuning the chemistry of the coatings by controlling parameters such as the plasma source or the feed gases (monomers and additives). For instance, gaseous precursors such as $C_2F_6$, $C_3F_8$, and $C_4F_8$ were studied in continuous wave, pulsed, and downstream radio-frequency plasmas.[7] The structure of the films deposited by continuous/wave (CW) plasma is highly cross-linked while pulsed plasma with long pulse-off times and afterglow plasma leads to the prevalence of the $CF_2$ component (i.e., less cross-linking). Favia et al.[5] obtained spectacular ribbon surface structures characterized by a strong $CF_2$ component, a small degree of crystallinity, and large water contact angles (WCA) higher than 1508 when TFE was deposited in modulated RF glow discharges at a low duty cycle. Hydrogen addition to saturated fluorocarbon monomers allows the deposition of coatings with variable F/C ratios and crosslinking degrees.[6]

The precursors used at atmospheric pressure are usually identical to those used in low-pressure processes (e.g., $CF_4$, $C_2F_4$, $C_2F_6$, $C_3F_8$, $C_3HF_7$, or $c—C_4F_8$), but the use of a carrier gas (usually helium or argon) is most often required when Working at atmospheric pressure. Although helium and argon have both demonstrated their efficiency in the plasma deposition of fluorocarbons, studies usually deal with the effects of a single gas, either argon or helium. One of the first investigations on the deposition of fluorocarbons by atmospheric plasmas was reported by Yokoyama et al,[8] working with a mixture of $C_2F_4$ and helium. The same mixture was used to modify the surface of poly(vinyl chloride) (PVC) and showed WCA of 100–110° but the modified surface was smoother that the pristine surface.[9] Fanelli et al.[10] demonstrated that atmospheric pressure dielectric barrier discharge (DBD) could also be useful in tailoring the coating properties while controlling the frequency or the addition of hydrogen. Varying the excitation frequency mainly affects the deposition rate, while the concentration ofH2 in the feed allows the control of the chemical composition of the films, additionally to the increase in the deposition rate (from 1 to 12nm.min$^{-1}$). Vinogradov et al.[11] reported the influence of additive gases such as $H_2$ and $O_2$ in a fluorocarbon-containing filamentary DBD. The presence of oxygen is known to shift the process from deposition to etching and induces then a decrease in the deposition rate.[2,12]

It is worth noticing that most of the precursors utilized in the plasma-enhanced chemical vapor deposition (PECVD) at low and atmospheric pressures are gaseous. Although it is not a new area of research at low pressure, only few studies have reported the use of fluorinated liquid solutions as precursors of the reaction.[13–15] Moreover, while some studies report the use of perfluorohexane, most of those do not report the use of pure $C_xF_y$ compounds and the highest WCAs obtained are 115°.[16–18] In the present study, the atmospheric pressure PECVD deposition and texturization of hydrophobic coatings using vapors from fluorinated liquid precursors at room temperature are investigated. We have studied the main effects of the nature of the carrier gas (argon and helium) through the characterizations of the plasma phase and the surface of the deposited layers. The filamentary/homogeneous character of the discharge is analyzed by a high-speed camera and electrical measurements. The influence of the carrier gas on the fragmentation of the precursor studied by mass spectrometry is discussed, along with the importance of the presence of impurities. The chemical composition, wettability, and morphology of the coatings are determined from X-ray photoelectron spectroscopy (XPS), WCA, and atomic force microscopy (AFM) measurements, respectively. Although a few studies have reported an influence of the carrier gas on the morphology of plasma-polymerized coatings, this work is, as far as we know, the first one aimed at studying such influence in the case of hydrophobic fluorinated films.[19,20] The effect of the carrier gas is indeed often neglected while it can be a key parameter for inducing an enhancement of the hydrophobic character of the coating. In that respect, argon and helium are used as carrier gases in the same plasma reactor in order to compare their effects on the coatings properties.

# 2. Experimental Section

## 2.1. Materials

The liquid precursorsnamelyperfluoro-2-methyl-2-pentene ($C_6F_{12}$) and perfluorohexane ($C_6F_{14}$) were provided by Fluorochem and were used without any further purification. Perfluorohexane is a mixture of two isomers, represented in Figure 1. Silicon Wafers (100) from Compart Technology Ltd. were used as substrates after being cleaned with methanol and isooctane. The glow discharge was sustained with argon (Air Liquide, ALPHAGAZ 1) or helium (Air Liquide, ALPHAGAZ 1).

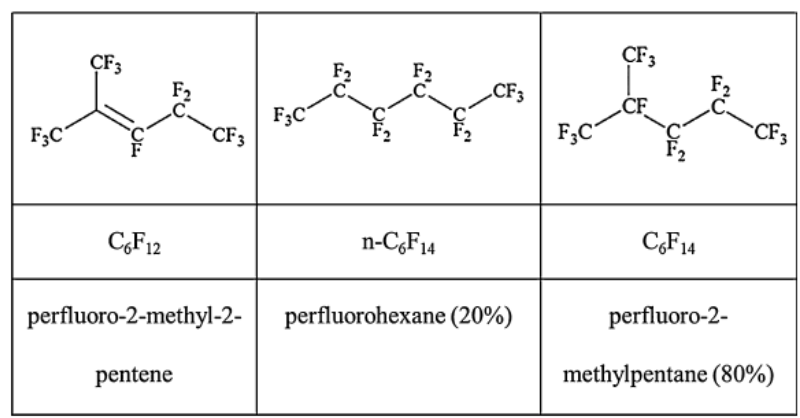

**Figure 1. Structures of the two liquid precursors $C_6F_{12}$ and the two $C_6F_{14}$ isomers.**

## 2.2. Plasma Polymerization

The plasma-polymerized (pp)-fluorinated films were deposited with a home-built dielectric barrier discharge (DBD) described in a previous study.[21] Both electrodes used were copper disks of 45mm diameter. The top electrode was covered with a 1.5mm thick Pyrex glass acting as a dielectric. The inter electrode gap was fixed at 3mm to avoid arcing at atmospheric pressure. To improve the film homogeneity, the grounded electrode was connected to a rotary system powered by a DC generator, thus allowing the rotation of the substrate at 60–80 rpm.

In order to prevent contamination, the chamber was first pumped down to a pressure of 2 torr. The atmospheric pressure was then reached by injecting the carrier gas (argon or helium). At that stage, the vapors of the precursor were introduced into the discharge by means of the carrier gas (second flow – 0.2 L.min⁻¹) bubbling in to the liquid, after being diluted with a primary flow of the carrier gas with a total carrier gas flow of5 L.min⁻¹. The precursor feed was estimated by measuring the mass variation of the bubbler for several flow rates. The injected amounts for 0.2 L.min ⁻¹ are as follows: Ar—$C_6F_{12}$ 782±44mg.min⁻¹, He—$C_6F_{12}$ 783±52mg.min⁻¹, Ar—$C_6F_{14}$ 768±84 mg.min⁻¹, and He—$C_6F_{14}$ 767±101mg.min⁻¹.

The operating frequency was set at 17.1 kHz with an output power set at 50W for the presented results, supplied by an AFSG10S (AFS entwicklungs- und vertriebs gmbh, Horgau, Germany) power generator. The deposition time varied from 30 to 360 s but most of the results presented correspond to films with a deposition time of 180 s for argon and 360 s for helium.

## 2.3. Electrical Measurements

The charge–voltage Lissajous method was used to determine the energy absorbed. A 10 nF capacitor ($C_{meas}$) was connected in series with the bottom electrode of the DBD device and the ground.[22] The voltages at the top electrode and across the capacitor were measured using two Tektronix voltage probes (P6015A and P6139B resp.) connected to a digital oscilloscope (Tektronix DPO 3032). The actual plasma power can be calculated by measuring the charge accumulation Q at the capacitor $C_{meas}$:

$$Q = C_{meas} * U_C$$

with $U_C$ the voltage across the capacitor. Plotting Q as a function of the voltage U at the top electrode gives a so-called Lissajous figure. The area enclosed by this diagram is the electrical energy consumed by the plasma per voltage cycle, which, multiplied by the operating frequency gives the actual plasma power.[22] The estimation of the absorbed energy is reported in Table 1 for each condition discussed in this article. The voltage and the current curves were measured by a high voltage probe (Tektronix P6015A) and a Rogowski coil (Pearson current monitor model 2877) both connected to a digital oscilloscope (Tektronix DPO 3032).

| Plasma condition | Absorbed energy (mJ) |
|---|---|
| Ar—$C_6F_{12}$ | $2.21 \pm 0.05$ |
| He—$C_6F_{12}$ | $2.36 \pm 0.08$ |
| Ar—$C_6F_{14}$ | $2.31 \pm 0.06$ |
| +Oxygen | $2.37 \pm 0.05$ |
| He—$C_6F_{14}$ | $2.42 \pm 0.09$ |
| +Oxygen | $2.49 \pm 0.08$ |

*Table 1. Estimation of the absorbed energy according to the Lissajous method (50 W, 0.2 L.min⁻¹ monomer/carrier gas, 5 L.min⁻¹ total carrier gas, 0.1 L min⁻¹ O₂ L.min⁻¹).*

## 2.4. High Speed Camera

The behavior of the DBD was followed via the brightness signal extracted from videos recorded with a Photron FastCam SA4 camera with an acquisition speed of 10 000–100 000 frames per second (fps). The profile lines are established by probing the brightness along a selected line of the video. The histograms are established by probing the brightness distribution over regions of interest by the Photron FastCam Viewer software. The level of brightness is expressed as a grayscale corresponding, in our case, to 256 levels of brightness.

Ar-plasma videos were acquired at 100 000 fps with a spatial definition of 192*128 pixels. Since the brightness of the helium plasma is much less intense, the videos were acquired with an acquisition speed of 10 000 fps to collect enough light for the shutter to acquire the image. It has to be noted that experiments for argon at 10 000 fps also present the same phenomenology (streamers visible), but with a higher brightness. All other parameters of acquisition were kept constant.

## 2.5. Mass Spectrometry (MS)

Mass spectrometry of the gas phase was performed with a Hiden analytical atmospheric gas analysis—QGA. The gases were collected through a PFA capillary located between the electrodes at the boundary of the plasma region. The secondary electron multiplier detector (SEM) was used to detect fragments with low partial pressures ($10^{-6}$–$10^{-13}$ Torr). In order to avoid excessive fragmentation of the precursor in the ionization chamber, the electron energy was set at 35 eV. The provided software MASsoft7 was used either to analyze the partial pressure as a function of the m/z ratio or to follow the partial pressure as a function of time for several specific m/z ratios, simultaneously.

## 2.6. Optical Emission Spectroscopy (OES)

Optical emission spectroscopy was performed with the SpectraPro-2500i spectrometer from ACTON research Corporation (0.500m focal length, triple grating imaging). The light emitted by the discharge was collected by an optical fiber and transmitted to the entrance slit (50mm) of the monochromator. Each optical emission spectrum was acquired with the 1 800 grooves.mm$^{-1}$ grating (blazed at 500 nm) and recorded over 10 accumulations with an exposure time of 50 ms. In order to tackle intensity variations as a function of the plasma conditions, the emissions of all the species were divided by the emission of the whole spectra (i.e., a continuum ranging from 250 to 850 nm).

## 2.7. Water Contact Angles (WCA)

A drop shape analyzer (Kruss DSA100) was used to measure static and dynamic water contact angles (WCA) onto the samples, according to the sessile drop method. The fitting method used is the ''Tangent 1.'' Advancing and receding contact angles were both measured by growing and shrinking the size of a single drop on the surface sample, from 0 to 15mL and back to 0mL at a rate of 30mL.min$^{-1}$. Advancing contact angles are the maximum angles observed during the droplet growth. The receding contact angles are measured by removing water from the droplet and are considered as the contact angles just before the contact surface reduction and distortion of the droplet.[23] All conditions lead to hydrophobic coatings characterized by an advancing WCA higher than 1158. The receding contact angles have been estimated from the dynamic measurements and the hysteresis seems to be constant whatever the advancing contact angle. A high hysteresis of about 20–30° would be indicative of a sticky surface characteristic of the Wenzel-type wetting mechanism and not the slippery Cassie–Baxter model.[24,25] For the sake of clarity, only the advancing WCA will be further discussed.

## 2.8. X-ray Photoelectron Spectroscopy

XPS analysis was performed on a Physical Electronics PHI-5600 photoelectron spectrometer. Survey scans were used to determine the elemental chemical composition of the surface. Narrow-region photoelectron spectra were used for the chemical study of the C 1s and F 1s. The spectra were acquired using the Mg anode (1253.6 eV) operating at 300 W. Wide surveys were acquired at a pass-energy of 187.5 eV with a five-scans accumulation (time/step: 50 ms, eV/step: 0.8) and high-resolution spectra of the C 1s peaks (and F 1s) were recorded at a pass-energy of 23.5 eV with an accumulation of 5 scans (time/step: 200 ms, eV/step: 0.05). The elementary composition was calculated after the removal of a Shirley background and using the sensitivity coefficients coming from the manufacturer's handbook: $S_C$=0.205, $S_F$=1, and $S_O$=0.63. The FWHM of the C 1s peak was around 2 with a constraint of 1.9–2.1 for the peak fitting.

## 2.9. Atomic Force Microscopy (AFM)

Atomic force microscopy (AFM) was used to analyze the surface morphology of the deposited films. AFM images were recorded in air with a Nanoscope IIIa (Veeco, Santa Barbara, CA) microscope operating in tapping mode. The probes were commercially available silicon tips with a spring constant of 24–52Nm$^{-1}$, a resonance frequency lying in the 264–339 kHz range, and a typical radius of curvature in the 5–10nm range. The images presented here are height images recorded with a sampling resolution of 512*512 data points and a scan size of 5*5 μm2.

## 2.10. Profilometry

The thickness measurements were performed with a stylus profiler Brücker dektak XT. The stylus with a 2mm radius scans the surface with a force of 1mg (0.01 mN) and the measurement was controlled and analyzed with the Vision 64 (Brüker, Billerica, MA) software. The thickness was estimated based on the height difference between the substrate surface and the surface of the coating. The height difference was obtained by placing an adhesive tape on the extremity of the substrate before the plasma deposition to maintain a non-coated area and by scratching the coating surface after deposition. The thickness value was evaluated by taking the average of each profile, and used to estimate the deposition rates.

# 3. Results and Discussion

## 3.1. Characterization of the Filamentary and Glow-Like Discharges

Depending on the conditions (gas, voltage, etc.), an atmospheric pressure dielectric barrier discharge can operate in a filamentary or glow mode.[26] In most cases, DBDs are non-uniform and consist of numerous microdischarges, with a lifetime of 10–100 ns and an electron density of $10^{14}–10^{15}$ cm$^{-3}$.[27–30] These micro-discharges (or streamers) are randomly distributed in the gas gap due to the dielectric material covering the electrodes. Glow (homogeneous) discharges are characterized by a uniform distribution of species constituting the plasma and electron densities of $10^{10}–10^{11}$ cm$^{-3}$.[31,32] Depending on the conditions, homogeneous and filamentary discharges can be obtained for a given carrier gas.[33] However, it is much more complex to form a homogeneous discharge with argon than helium as illustrated in Figure 2 for $C_6F_{12}$ polymerized in the DBD source.

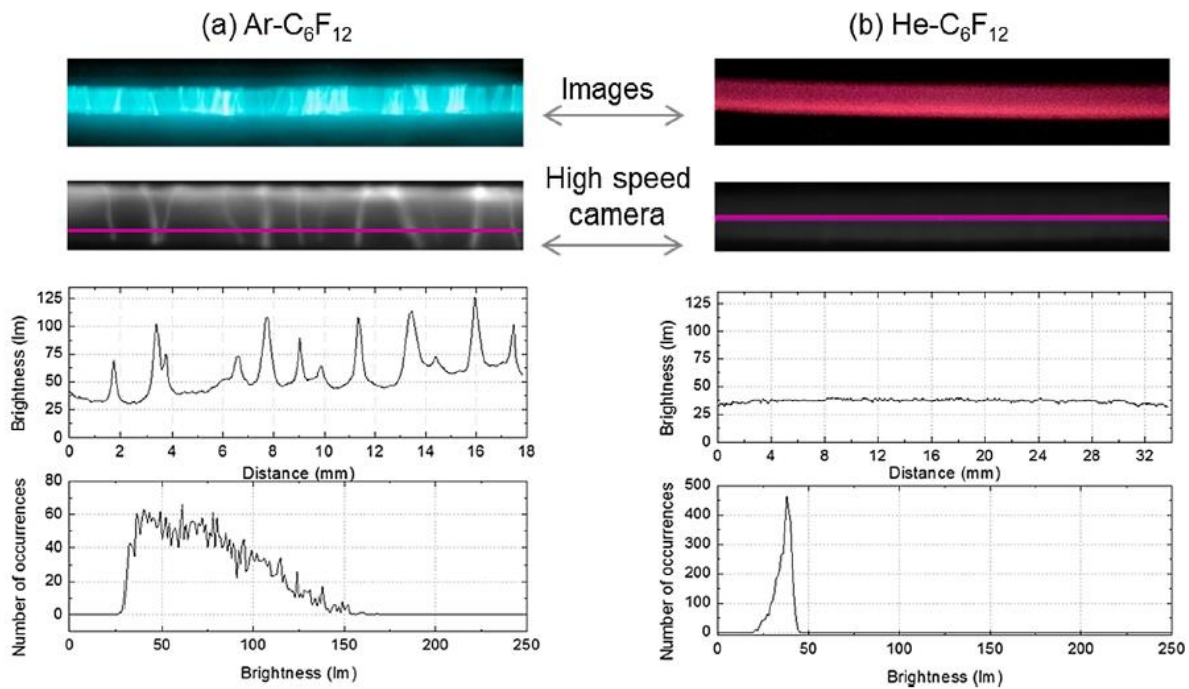

**Figure 2. Behavior of the discharge as a function of the carrier gas (a) Ar—$C_6F_{12}$ (b) He—$C_6F_{12}$ (50 W, 0.2 L.min$^{-1}$ monomer/carrier gas, 5 L.min$^{-1}$ total carrier gas), including pictures of the discharges, brightness profiles versus distance, and histograms reporting number of occurrences versus brightness (obtained from a high-speed camera—acquisition time: Ar: 1/100 000 s, He: 1/10 000 s).**

The presence of micro-discharges inducing the filamentary character of the argon discharge compared to the more homogeneous helium discharge is clearly observed in Figure 2. The brightness profile has been plotted along a line represented by the purple segment on the pictures. Along these lines, the streamers are made visible by peaks of higher brightness. In the case of Ar—$C_6F_{12}$, several peaks can be observed which are the steamers gathered on a single picture extracted from longer times of acquisition. This picture is representative of the behavior observed in several videos. However, the number of streamers remains approximately constant over the time of the whole video. In the case of He—$C_6F_{12}$, the one dimensional brightness line extracted from the picture is rather different: the brightness remains constant and stable without any peak, highlighting the homogeneity of the discharge. As in the previous case, the profile line is similar for longer acquisition times. The histograms of Figure 2 represent the brightness distribution over a larger region of interest which comprises the entire region where the plasma is developed. In the He—$C_6F_{12}$ plasma, all the events in the discharge are similar, which leads to a rather homogeneous distribution of the brightness centered at a low level of brightness (<50 luma). Conversely, the histogram of brightness of the Ar—$C_6F_{12}$ plasma shows a broader distribution, which indicates the occurrence of events such as the presence of streamers involving events with higher brightness. It has to be noted that the mean intensity of brightness is much higher in the case of Ar-plasma than in the case of He-plasma.

As a supplement to the visual characterization of the discharge, the current curves presented in Figure 3 evidence the highly filamentary character of the argon discharge compared to helium. Moreover, a blue color discharge is observed only for argon plasmas. Although rarely discussed in the literature, this color seems to be characteristic of $CF_2$ emission and would be the signature of a higher fragmentation, higher $CF_2$ concentration in the films, and higher deposition rate recorded for argon plasmas.[34] It has to be noted that these observations are similar for the $C_6F_{14}$ precursor.

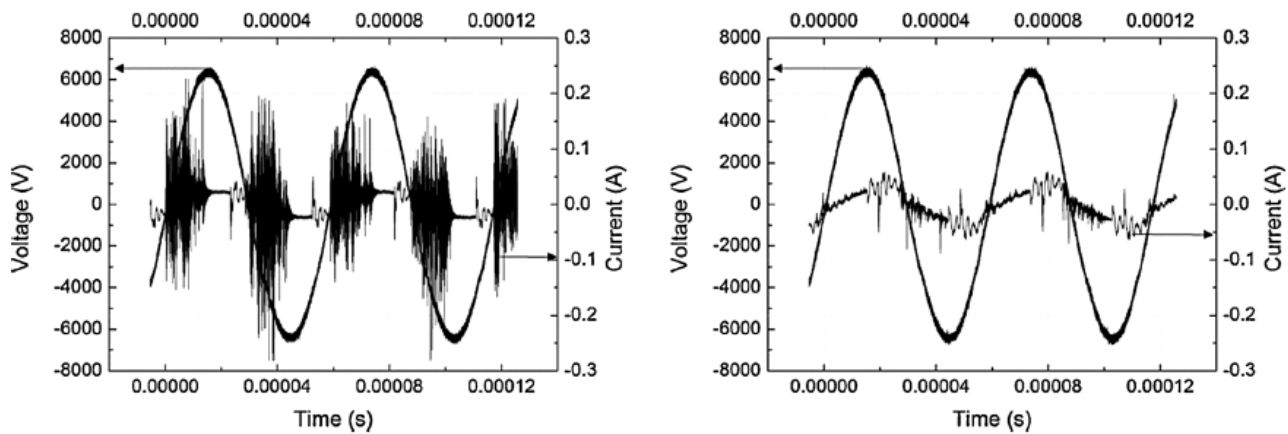

*Figure 3. Voltage and current curves of (a) Ar—$C_6F_{12}$ and (b) He—$C_6F_{12}$ plasma (50W, 0.2 L.min⁻¹ monomer/carrier gas, 5 L.min⁻¹ total carrier gas).*

## 3.2. Characterization of the Plasma Phase

In order to evaluate the fragmentation in argon and helium plasmas, mass spectra as a function of the m/z ratio have been plotted to identify the interesting fragments and then, their intensity has been studied as a function of time. The mass spectrum of the $C_6F_{14}$ precursor is presented in Figure 4(a), and is similar to the one referenced by the National Institute of Standards and Technology.[35] Both measured and reference spectra have been considered to define the most probable fragments released from the fragmentation in the mass spectrometer. The considered spectrum was measured in an argon atmosphere but the fragments are identical in the case of helium and differ only in intensity. For the sake of clarity and to focus on $C_xF_y$ fragments, we only present m/z higher than 50.

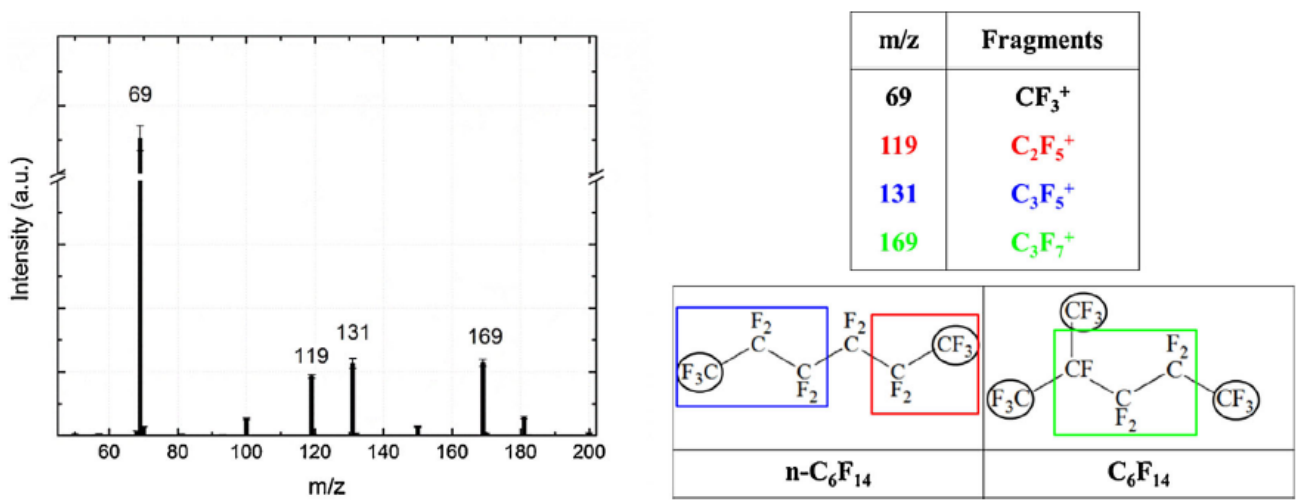

**Figure 4.** (a) Mass spectrum of $Ar—C_6F_{14}$ (50 W, 0.2 L.min$^{-1}$ monomer/carrier gas, 5 L.min$^{-1}$ total carrier gas), (b) List and structures of the most probable fragments according to the mass spectrum and the literature.

Considering the structure of the precursors in Figure 4(b), the fragments released from the mass spectrometer are assumed to be direct products of the main fragmentation of the precursors. For the perfluorohexane, the $C_3F_7^+$ fragment arises from the symmetric rupture of the $n—C_6F_{14}$. The formation of $CF_3^+$ occurs by fragmentation of the end of both configurations, so does the formation of $C_2F_5^+$. The presence of $C_3F_5^+$ fragments can be explained, for example, by the release of the $CF_3^+$ groups in the isomer structure of the molecule.

In order to observe the direct effect of the plasma on the fragmentation, we monitored all the selected m/z ratios as a function of time, and laid out the most relevant variations. The period before 800 s corresponds to the gas/precursor flowing in the reactor before plasma ignition, and the intensity of the corresponding monitored fragment is referred to as the baseline. The plasma was then turned on and the behavior of the partial pressure of the fragments was followed for about 5 min, before switching off the plasma again. At that point, the precursor was still injected in the reactor until the intensity returned back to the baseline.

First, it should be mentioned that mass spectrometry is not a direct method to describe chemistry kinetics in the discharge because species produced in the plasma must travel throughout all the spectrometer capillary and the ionization chamber of the mass spectrometer induces a second fragmentation. However, interesting information can still be extracted. Indeed, in the argon plasma, the intensity of three of the main fragments ($CF_3^+$, $C_3F_5^+$, and $C_3F_7^+$) coming from the fragmentation of $C_6F_{14}$ in the mass spectrometer decreases, while the intensity of many other (small) fragments such as CF, $CF_2$, $C_2F_4$, or $C_3F_3$ (not presented here) increases. In helium, the effect of the plasma is less pronounced, but all the intensities of the detected fragments are increased when the discharge is turned on, as shown by the dashed curves in Figure 5. We assume that these observations can be interpreted as a sign of higher fragmentation in argon. Despite the complexity of the reactions in the discharge, the differences between the two systems (''mass spectrometry'' vs. ''plasma+mass spectrometry'') can be briefly summarized in the next scheme (Figure 6), where only the main detected fragments are discussed.

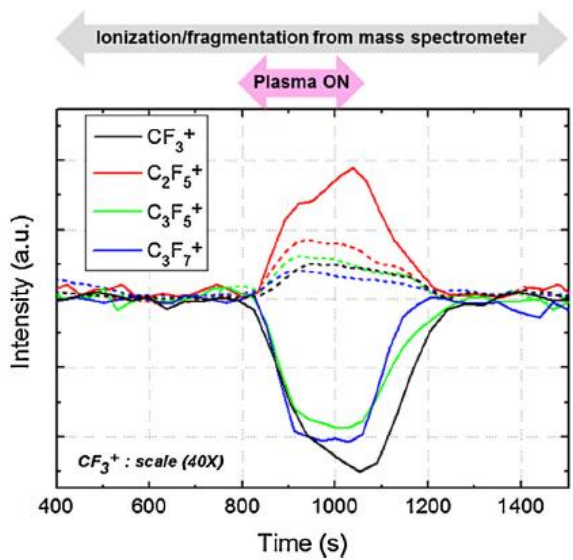

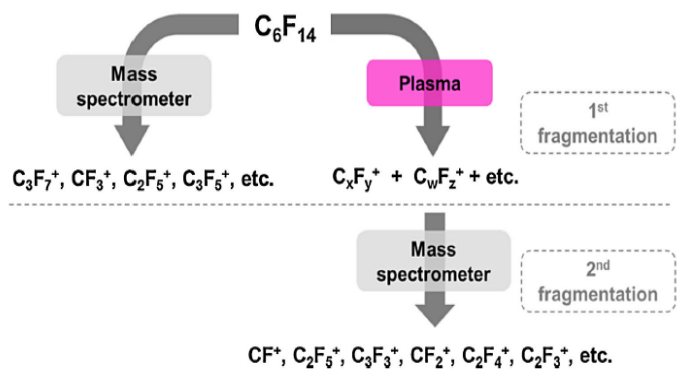

Figure 5. Intensity of the $CF_3^+$, $C_2F_5^+$, $C_3F_5^+$, and $C_3F_7^+$ fragments generated during the APPECVD of $C_6F_{14}$ as a function of time, in helium (- - - dashed curves) and argon (—— full curves) (50 W, 0.2 L.min$^{-1}$ monomer/carrier gas, 5 L.min$^{-1}$ total carrier gas).

Figure 6. Scheme of the possible fragmentation of $C_6F_{14}$ (left) in the mass spectrometer and (right) in the plasma, followed by the mass spectrometer ($C_xF_y^+$ and $C_wF_z^+$ assuming to be $CF_3^+$, $C_3F_7^+$, and C3F5+ according to mass spectrometry results).

As the fragmentation of the precursor is carried out upstream of the mass spectrometer, some of the $C_3F_7^+$, $CF_3^+$ and $C_3F_5^+$ ions are already present in the discharge and would be even more fragmented by the mass spectrometer, explaining their decreasing intensity. We suggest that the reason why the $C_2F_5^+$ fragment is not reduced is because it is a possible product of the fragmentation of $C_3F_7^+$. Indeed, $C_3F_7^+$ can lead to $CF_3^+$, $C_2F_4^+$, $CF_2^+$, $C_2F_5^+$; possible products from $C_3F_5^+$ are $CF^+$, $C_2F_4^+$, $CF_2^+$, $C_2F_3^+$, and $CF_3^+$ can provide $CF_2^+$ or $CF^+$ with a release of fluorine atoms which would mostly be detected as negative charged species. Unfortunately, negative spectra cannot be obtained from our mass spectrometer.

According to Figure 5, the consumption of the most intense fragments did not occur in helium plasma. The interaction between the molecules of the precursor and the active species of the plasma is supposed to be quite distinct compared to argon plasma. As previously mentioned, the electrons and micro-discharges densities in helium are much lower and could then explain the lower fragmentation.

Moreover, helium is known to possess excited states that are very efficient to excite all the impurities present between the electrodes.[32] The carrier gas can indeed be excited into electronic excited metastable states which induce Penning ionization of other species if the energetic level of the metastable is higher than the ionization potential of that species.[36] In the case of helium, the metastable atoms are energetic enough ($1s^2 s^3 S_1$ and $1s^2 s^1 S_0$ states with energies of 19.82 and 20.62 eV, respectively) to ionize most atoms and molecules, explaining this efficiency to excite impurities. This is in contrast with the argon metastable atoms which have energies (11.55 and 11.72 eV for $3p^5 4s^3 P_2$ and $3p^5 4s^3 P_0$ states, respectively) lower than the ionization potential of $O_2$ (12.07 eV) or $N_2$ (15.6 eV).[32,36]

The efficiency of generating atomic oxygen in helium plasmas is illustrated in Figure 7 which represents the emission intensity of the line at 777 nm. Indeed, adding $O_2$ to helium plasmas induces a strong increase in the atomic oxygen emission while small variations are observed in the argon discharges. The involvement of oxygen might be crucial for the plasma polymerization, and oxygen-containing species are an important part of the products detected in helium plasmas. The helium excited states seem indeed to activate oxygen impurities much more easily than argon since products such as $CO_2^+$, $COF^+$, or $COF_2^+$ were strongly detected by mass spectrometry (see Figure 8).

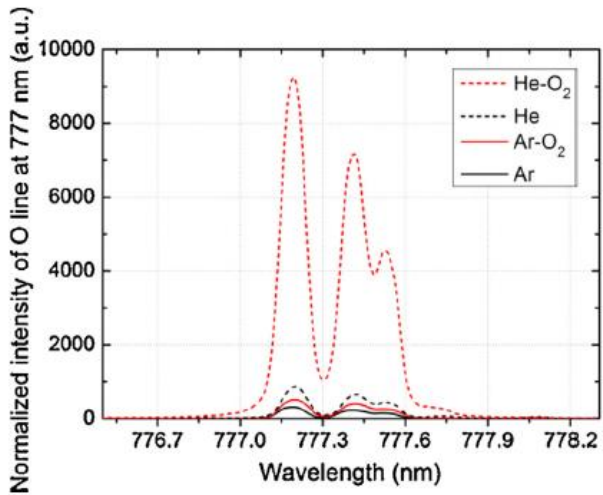

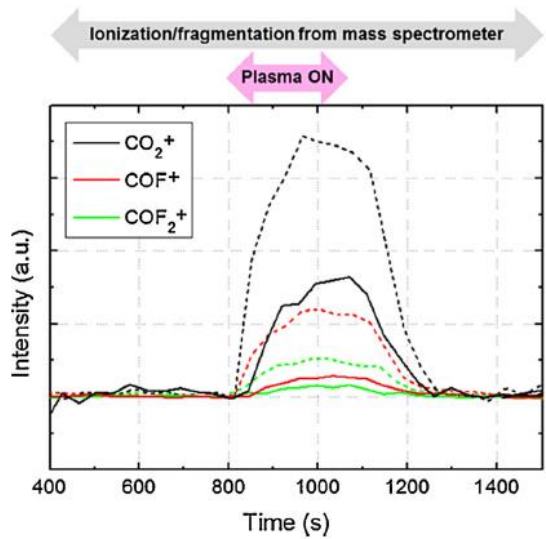

*Figure 7. Emission intensity of atomic oxygen at 777 nm as a function of the carrier gas and the addition of molecular oxygen (50 W, 0.2 L.min⁻¹ monomer/carrier gas, 5 L.min⁻¹ total carrier gas, 0.1 L.min⁻¹ O₂ L.min⁻¹).*

*Figure 8. Intensity of the $CO_2^+$, $COF^+$, and $COF_2^+$ fragments generated during the APPECVD of $C_6F_{14}$ as a function of time, in helium (- - - dashed curves) and argon (—— full curves) (50 W, 0.2 L.min⁻¹ monomer/carrier gas, 5 L.min⁻¹ total carrier gas).*

We mostly discuss the case of $C_6F_{14}$ as, according to the literature, all its main fragments have their m/z ratios lower than 200 and can thus be detected.[35] The case of $C_6F_{12}$ is more complicated to interpret because all its main fragments are not observable with our mass spectrometer ($C_5F_9^+$ and $C_6F_{11}^+$ at m/z of 231 and 281, respectively) and because of the presence of high reactive C=C double bonds. The intensity of $C_4F_7^+$ at 181 is decreasing while the intensity of small fragments is increasing. This could be explained either by the fragmentation of the precursor or by the oligomer formation in the gas phase as suggested by Nisol et al.[37] and the higher deposition rate detailed in the next section. However, a better interpretation of the $C_6F_{12}$ behavior requires information of higher m/z ratios in order to analyze a possible polymerization.

## 3.3. Surface Properties of the Deposited Coatings

### 3.3.1. Deposition Rates

The comparison of the deposition rates shows that both the precursors and the carrier gases induce a different reactivity, as presented in Table 2. Much faster deposition is obtained with the $C_6F_{12}$ precursor, which is assumed to be related to the higher reactivity of the C=C double bond present in the molecule. This behavior was also highlighted in the case of hydrocarbon or acrylate molecules, among others.[38–40] Moreover, Yasuda et al.[39] stated that the plasma polymerization of unsaturated compounds proceeds by the opening of double or triple bonds. If this interpretation is correct, most of the energy would then be consumed to break the double bond for polymerization. A strong drop in the deposition rate is also observed when helium is used instead of argon, although the same amount of the precursor is injected into the discharge. This can be related to the observation by Jiang et al.[41] that a higher amount of activated precursor is produced along the streamers. The streamers hitting the surface cause a local heating and an activation of the surface which induces a higher growth rate at these locations.[41] The streamers being randomly distributed, they move and cover the entire substrate surface.

| Deposition rate (nm min$^{-1}$) | $C_6F_{12}$ | $C_6F_{14}$ |
|---|---|---|
| Argon | $240 \pm 20$ | $50 \pm 5$ |
| Helium | $25 \pm 3$ | $< 5$ |

*Table 2. Average deposition rates of $C_6F_{12}$ and $C_6F_{14}$ in argon and helium (50 W, 0.2 L.min$^{-1}$ monomer/carrier gas, 5 L.min$^{-1}$ total carrier gas).*

### 3.3.2. Chemical Composition of the Coatings

The chemical composition of the fluorinated coatings has been characterized by XPS. According to the literature, $C_xF_y$ films deposited by plasma are usually characterized by a 5-component curve fitting of the XPS C 1s high-resolution peak.[10,11,42,43] The fitting of the C1s peak of pp—$C_6F_{12}$ films was therefore achieved with CF$_3$ (294.2±0.2 eV), CF$_2$ (292.2±0.2 eV), CF (289.9±0.2 eV), CCF (287.8±0.2 eV), and CC (284.9±0.2 eV) components; with the CCF being a C—C bond shifted to higher binding energy because of the electronegative environment. The curves are represented in Figure 9.

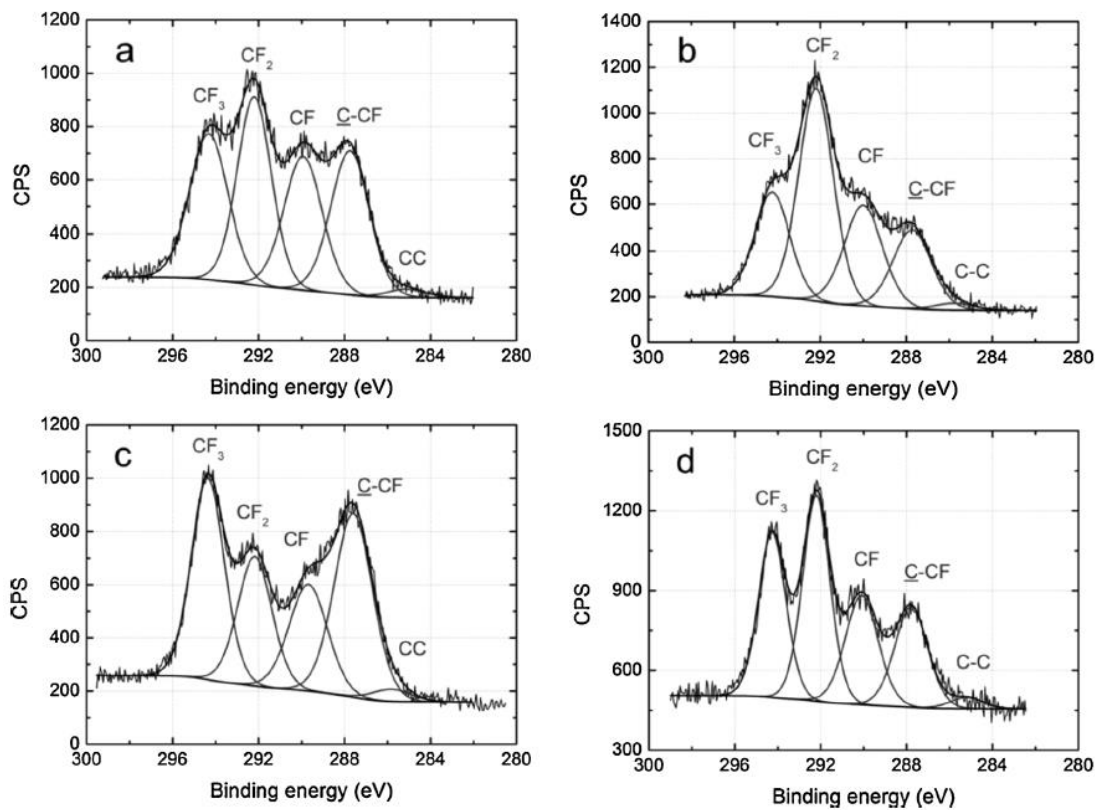

*Figure 9. XPS spectra of (a) Ar—$C_6F_{12}$, (b) Ar—$C_6F_{14}$, (c) He—$C_6F_{12}$, and (d) He—$C_6F_{14}$ plasma-polymerized films (50 W, 0.2 L.min$^{-1}$ monomer/carrier gas, 5 L.min$^{-1}$ total carrier gas).*

The F/C ratio of the $C_6F_{12}$ and $C_6F_{14}$ precursors is 2 and 2.3, respectively. In the plasma-polymerized films, the ratio decreases down to 1.17 and 1.32, respectively, in argon and down to 1.11 and 1.37, respectively, in helium. This partial defluorination of the precursor explains the presence of branched components such as CCF and CC, as evidenced in the C 1s high-resolution peak. Despite this observation, the much more intense percentage of $CF_2$ in the pp—$C_6F_{14}$ (e.g., $CF_2$ is 10% higher in the pp—$C_6F_{14}$/Ar than in the pp—$C_6F_{12}$/Ar film as seen in Table 3) seems to indicate that the signature of the precursor is partly preserved in the coating. The detection of the silicon substrate for the He—$C_6F_{14}$ coatings indicates that the deposited layer is very thin and confirms the very low deposition rate previously reported. We will not detail this spectrum further as the contribution from the substrate might induce errors in the interpretation.

| Composition (%) | F | O | C | Si | $CF_3$ | $CF_2$ | CF | CCF | CC |
|---|---|---|---|---|---|---|---|---|---|
| Ar—$C_6F_{12}$ | 53.9 ± 0.9 | 0 | 46.1 ± 0.9 | 0 | 23.4 ± 2.0 | 28.8 ± 1.1 | 22.0 ± 0.3 | 23.7 ± 1.3 | 2.2 ± 0.7 |
| He—$C_6F_{12}$ | 51.4 ± 0.9 | 2.1 ± 0.3 | 46.6 ± 0.4 | 0 | 29.5 ± 0.7 | 20.6 ± 0.7 | 18.5 ± 0.4 | 29.3 ± 1.3 | 2.0 ± 0.4 |
| Ar—$C_6F_{14}$ | 56.9 ± 0.7 | 0 | 43.1 ± 0.7 | 0 | 20.9 ± 0.9 | 39.8 ± 1.4 | 21.3 ± 1.0 | 16.3 ± 1.9 | 1.8 ± 0.4 |
| He—$C_6F_{14}$ | 54.9 ± 0.9 | 1.5 ± 0.6 | 39.9 ± 1.5 | 3.5 ± 1.4 | 24.6 ± 0.9 | 33.7 ± 1.7 | 20.8 ± 0.9 | 18.0 ± 1.4 | 2.8 ± 1.3 |

*Table 3. Surface composition of pp—$C_xF_y$ films as a function of the precursor and the carrier gas (50 W, 0.2 L.min⁻¹ monomer/carrier gas, 5 L.min⁻¹ total carrier gas).*

Additionally to the precursor, the choice of the carrier gas influences the surface chemical composition of the coating. The composition of the films polymerized in helium is indeed rather different from those polymerized in argon (Figure 9). In a helium plasma, the pp—$C_6F_{12}$ coatings have higher concentrations of $CF_3$ and CCF groups, for identical plasma conditions. Given the structure of the precursor (Figure 1), as well as the lower electron density, the lower electron energy and the lower micro-discharges density of the more homogeneous (glow-like) helium discharge, we could therefore assume that the precursor structure is more preserved when using helium. Moreover, it has to be noted that a small incorporation of oxygen (<3%) was observed only with helium. Helium has been shown to be reactive toward oxygen impurities.[31] This property is important as it can explain why small amounts of oxygen are present in the film polymerized in helium, whatever the experimental conditions and why energetic helium species are less involved in the fragmentation/polymerization of the monomer (although the fragmentation is assumed to be performed mainly by electron impact). These species can indeed be consumed by the impurities.

### 3.3.3. Hydrophobicity and Morphology of the Coatings

According to Figure 10, both the carrier gas and the precursor have a significant impact on the morphology of the coating, thus inducing variations in the hydrophobicity. In helium plasmas, the water contact angle and the roughness are below 120° and 2 nm, respectively, which indicates the presence of a smooth, low-energy surface. This is not surprising due to the more homogeneous discharge and stable WCAs recorded as a function of the treatment time in the case of tetrafluoromethane deposition in helium at atmospheric pressure.[8]

The argon-deposited coatings are very different since the roughness required to increase the hydrophobicity can be obtained by the filamentary character of those discharges. Advancing water contact angles as high as 140° are then reached. In contrast with helium plasmas, the precursor plays a role in the morphology of the coatings. For a similar thickness of approximately 300nm, Ar—$C_6F_{14}$ discharges induce not only a higher roughness and hydrophobicity but also a roughness structure totally different from the $C_6F_{12}$ precursor as shown in Figure 11. The alveolar structures of approximately 100 nm are quite regular and could be characteristic of the filamentary discharge inducing an oriented etching of the surface in competition with the polymerization. The presence of the C=C double bond in $C_6F_{12}$ and its lower F/C ratio could promote the polymerization character instead of the etching, as it is the case for $C_6F_{14}$.

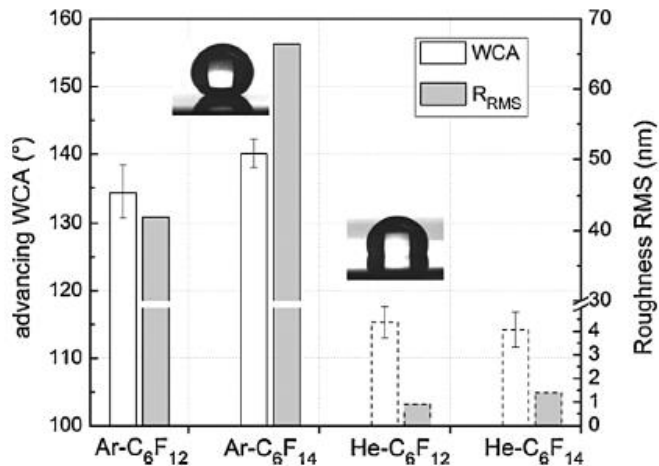

*Figure 10. Advancing WCA and Roughness RMS of the plasma polymerized films. Illustration of the drop behavior of (left) Ar—$C_6F_{14}$ and (right) He—$C_6F_{12}$ coatings (50 W, 0.2 L.min⁻¹ monomer/carrier gas, 5 L.min⁻¹ total carrier gas).*

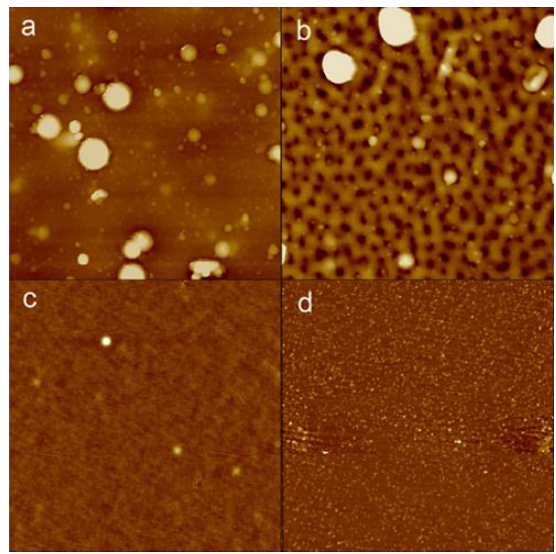

*Figure 11. 5*5 μm² AFM images of (a) Ar—$C_6F_{12}$, (b) Ar—$C_6F_{14}$, (c) He—$C_6F_{12}$, and (d) He—$C_6F_{14}$ plasma-polymerized films (50 W, 0.2 L.min⁻¹ monomer/carrier gas, 5 L.min⁻¹ total carrier gas). The vertical color scale corresponds to 300nm for images a and b, and to 30 nm for images c and d.*

## 4. Conclusion

The plasma polymerization deposition of fluorinated coatings was investigated as a function of the carrier gas (argon and helium), in the case of the precursors perfluoro-2-methyl-2-pentene ($C_6F_{12}$) and perfluorohexane ($C_6F_{14}$). We showed that helium and argon plasmas react differently toward the fluorocarbon precursors. The nature of the carrier gas induces a drastic effect on the chemical composition, morphology, and therefore hydrophobicity of the films and has been, as far as we know, never discussed in terms of hydrophobic fluorocarbons. These differences are evaluated and are assumed to come from the generation of both glow-like discharges (in helium) versus filamentary discharges (higher electron energies and densities) in argon. We showed that hydrophobic surfaces with water contact angles (WCA) higher than 115° were obtained only in the presence of argon. This is because of the filamentary argon discharge induces higher deposition rates and rougher coatings than the more homogeneous helium discharge. The formation of activated sites on the growing film and the activated monomeric species is much more important for the argon plasma. Moreover, the easier activation of impurities by helium metastable states seems to lead to the presence of a small percentage of oxygen in the coating, a lower deposition rate and a weaker fragmentation/polymerization of the monomer.

In addition to the surface characterization, mass spectrometry and optical emission spectroscopy measurements were performed in order to describe the gas phase. Both techniques allowed identifying the presence of oxygen and its involvement in the reactions taking place in helium plasmas. The analysis of fragments issued from the mass spectrometry while the plasma is ignited seems to confirm the higher fragmentation in argon. However, the limitation in the m/z ratio of the apparatus prevents us from analysing fragments resulting from a possible polymerization in the gas phase. Supplementary information from IR in the gas phase would then be very useful, since by-products such as $COF_2$ or $COF$ could be detected in addition to products originating from the polymerization.

## 5. Acknowledgements

This work was part of the IAP (Interuniversity Attraction Pole) programs ''PSI—Physical Chemistry of Plasma Surface Interactions'' and ''Functional Supramolecular Systems'' financially supported by the Belgian Federal Office for Science Policy (BELSPO). This work was also financially supported by the FNRS (Belgian National Fund for Scientific Research) for an ''Aspirant Grant (FRFC Grant No. 2.4543.04)''. CB thanks the Fonds de la Recherche Scientifique (F.R.S.-FNRS) for financial support (PhD grant), TVdB and CB also thank the Wallonia-Brussels Federation (Action de Recherches Concertees n8AUWB 2010-2015/ ULB15) for the acquisition of equipment. Research in Mons is also supported by the Region Wallonne/European Commission FEDER program (SMARTFILM project) and FNRS-FRFC. The authors want to thank Dr. Jennifer Christophe for the colored pictures of the discharges.